# Inhomogeneous Magnetoelectric Effect on Defect in Multiferroic Material: Symmetry Prediction

**B Tanygin**[1]

Taras Shevchenko Kyiv National University, Radiophysics Faculty, Kyiv, Ukraine

E-mail: b.m.tanygin@gmail.com

**Abstract.** Inhomogeneous magnetoelectric effect in magnetization distribution heterogeneities (0-degree domain walls) appeared on crystal lattice defect of the multiferroic material has been investigated. Magnetic symmetry based predictions of kind of electrical polarization distribution in their volumes were used. It was found that magnetization distribution heterogeneity with any symmetry produces electrical polarization. Results were systemized in scope of micromagnetic structure chirality. It was shown that all 0-degree domain walls with time-noninvariant chirality have identical type of spatial distribution of the magnetization and polarization.

## 1. Introduction
Coupling mechanism between magnetic and electric subsystem in multiferroics [1] is of considerable interest for fundamentals of condensed matter physics as well as for their application in the novel multifunctional devices [2]. Magnetically induced ferroelectricity is caused by homogeneous [3-6] or inhomogeneous [7-10] magnetoelectric interaction. The last appears in spiral multiferroics [11] or micromagnetic structures like domain walls (DWs) [7,12-17] including cases of magnetization distribution heterogeneities (0-degree DWs). Ferroelectricity developed from micromagnetic structure can appear in any magnetic material even in centrosymmetric one [18]. It was shown that inhomogeneous magnetoelectric effect is closely related to the magnetic symmetry of the system that had been demonstrated for specific cases of DWs [12]. Building of symmetry classification of 0°-DWs leads to determination of the type of electric polarization rotation in volume of any magnetic 0°-DW. Building of complete symmetry classification of one-dimensional magnetic heterogeneities and qualitative description of induced ferroelectricity in their volume is the aim of this report.

---

[1] B.M. Tanygin, 64 Vladimirskaya str., Taras Shevchenko Kyiv National University, Radiophysics Faculty. MSP 01601, Kyiv, Ukraine.

## 2. Magnetoelectric effect on crystal lattice defect

Let us describe the magnetoelectric effect in the magnetization distribution heterogeneity (i.e. 0°-DW). Such micromagnetic structure appears on the crystal lattice defect according to the localized change of the magnetocrystalline anisotropy. Most general expression for the free energy $F_{ME}$ of the magnetoelectric interaction is produced from symmetry analysis. All following analysis is phenomenological (no specific types of the microscopic interactions were considered). According to the Neumann's principle, the $F_{ME}$ terms must be the invariant of the crystal crystallographic class $G_P^\infty$ (magnetic point symmetry of paramagnetic phase). The well known homogeneous linear magnetoelectric coupling term is given by [3-6]:

$$F_{ME}^{(00)} = f_{i,ab}^{(00)} P_i M_a M_b \quad (1)$$

For the case of the polarization produced by the inhomogeneous magnetization it should be supplemented by the terms with magnetization derivates [7,12]:

$$F_{ME}^{(10)} = f_{ik,ab}^{(10)} P_i M_a \partial M_b / \partial x_k \quad (2)$$

The produced polarization distribution is inhomogeneous as well. Consequently, the following terms can be introduced:

$$F_{ME}^{(01)} = f_{il,ab}^{(01)} M_a M_b \partial P_i / \partial x_l \quad (3)$$

$$F_{ME}^{(11)} = f_{ikl,ab}^{(11)} M_a (\partial M_b / \partial x_k) \partial P_i / \partial x_l \quad (4)$$

The terms with higher derivates are next in order of magnitude [12]. The structures of the tensors $\hat{f}^{(00)}$, $\hat{f}^{(10)}$, $\hat{f}^{(01)}$, and $\hat{f}^{(11)}$ are defined by the crystal symmetry. The terms (1-4) relates to the local (short-range) magnetoelectric interactions [7,12]. The non-local interactions between magnetization and electric polarization distributions can be coupling via the spatially distributed stress tensor [19]. The free energy term $F_{non-loc}\{M(r), P(r)\}$ of the non-local interactions is not the invariant of the crystal crystallographic class $G_P^\infty$ in general case. This peculiarity of the Neumann's principle relates to the fact that crystallographic class does not completely describe symmetry of the medium. Symmetry of the crystal surface (crystal shape with magnetic point group $G_S$) restricts the symmetry of the medium in case of the non-local interactions which strongly depends on the boundary conditions. Consequently, the actual magnetic point symmetry of paramagnetic phase for non-local interactions is [20]:

$$G_P = (G_P^\infty \cap G_S) \subseteq G_P^\infty \quad (5)$$

The term $F_{non-loc}\{M(r), P(r)\}$ should be invariant of the group (5). Total free energy describing the magnetoelectric coupling in the 0°-DW is determined by the:

$$F_{ME} = F_{ME}^{(00)} + F_{ME}^{(10)} + F_{ME}^{(01)} + F_{ME}^{(11)} + F_{non-loc}\{M(r), P(r)\} \quad (6)$$

Solving the variational problem using the free energy term (6) allows obtaining of the function $P(r)$ produced by the magnetoelectric effect as it was shown in [21]. In the case of the one-dimensional model (planar DW) this function $P(z)$ can be described based on the symmetry analysis only [7,12]. Here Z axis is directed along the DW plane normal. It was shown [22] that there are 42 magnetic point groups of 0°-DW (table 1-3) which were systemized depend on their chirality (based on the novel chirality definition [23-26]). Specific component type (A) labels the odd function, (S) labels the even function and (A,S) labels the sum of odd and even ones. Relation between the magnetic point group of the DW and these order parameter component types is described in [22].

**Table 1.** Types of spatial distribution of electric polarization induced by the inhomogeneous magnetoelectric effect in volume of magnetic 0°-DWs with the time-invariant chirality.

| Magnetic point group | $M_x(z)$ | $M_y(z)$ | $M_z(z)$ | $P_x(z)$ | $P_y(z)$ | $P_z(z)$ |
|---|---|---|---|---|---|---|
| $2'_x 2_y 2'_z$ | A | S | 0 | 0 | 0 | A |
| $2'_z$ | A,S | A,S | 0 | 0 | 0 | A,S |
| $2'_x$ | A | S | S | S | A | A |
| $2_y$ | A | S | A | A | S | A |
| 1 | A,S | A,S | A,S | A,S | A,S | A,S |
| $2_z$ | 0 | 0 | A,S | 0 | 0 | A,S |
| $3_z$ | 0 | 0 | A,S | 0 | 0 | A,S |
| $4_z$ | 0 | 0 | A,S | 0 | 0 | A,S |
| $6_z$ | 0 | 0 | A,S | 0 | 0 | A,S |
| $2_z 2'_x 2'_y$ | 0 | 0 | S | 0 | 0 | A |
| $3_z 2'_x$ | 0 | 0 | S | 0 | 0 | A |
| $4_z 2'_x 2'_y$ | 0 | 0 | S | 0 | 0 | A |
| $6_z 2'_x 2'_y$ | 0 | 0 | S | 0 | 0 | A |

**Table 2.** Types of spatial distribution of electric polarization induced by the inhomogeneous magnetoelectric effect in volume of magnetic 0°-DWs with the time-noninvariant chirality.

| Magnetic point group | $M_x(z)$ | $M_y(z)$ | $M_z(z)$ | $P_x(z)$ | $P_y(z)$ | $P_z(z)$ |
|---|---|---|---|---|---|---|
| $m'_x$ | 0 | A,S | A,S | 0 | A,S | A,S |
| $m'_x m'_z 2_y$ | 0 | S | A | 0 | S | A |
| $m'_z$ | S | S | A | S | S | A |
| $m'_x m'_y 2_z$ | 0 | 0 | A,S | 0 | 0 | A,S |
| $3_z m'_x$ | 0 | 0 | A,S | 0 | 0 | A,S |
| $4_z m'_x m'_{xy}$ | 0 | 0 | A,S | 0 | 0 | A,S |
| $6_z m'_x m'_y$ | 0 | 0 | A,S | 0 | 0 | A,S |

**Table 3.** Types of spatial distribution of electric polarization induced by the inhomogeneous magnetoelectric effect in volume of magnetic achiral 0°-DWs.

| Magnetic point group | $M_x(z)$ | $M_y(z)$ | $M_z(z)$ | $P_x(z)$ | $P_y(z)$ | $P_z(z)$ |
|---|---|---|---|---|---|---|
| $m'_y m_x 2'_z$ | A,S | 0 | 0 | 0 | 0 | A,S |
| $m_y$ | A,S | 0 | 0 | 0 | A,S | A,S |
| $m_z m'_y 2'_x$ | A | 0 | S | S | 0 | A |
| $m_z$ | A | A | S | S | S | A |
| $m_y m'_x m'_z$ | 0 | S | 0 | 0 | 0 | A |
| $m_y m'_z 2'_x$ | 0 | S | 0 | S | 0 | A |
| $2_y/m_y$ | 0 | S | 0 | A | 0 | A |
| $2'_z/m'_z$ | S | S | 0 | 0 | 0 | A |
| $2'_x/m'_x$ | 0 | S | S | 0 | A | A |
| $\bar{1}$ | S | S | S | A | A | A |
| $2_z/m_z$ | 0 | 0 | S | 0 | 0 | A |
| $m_z m'_x m'_y$ | 0 | 0 | S | 0 | 0 | A |
| $\bar{6}_z$ | 0 | 0 | S | 0 | 0 | A |
| $\bar{6}_z m'_y 2'_x$ | 0 | 0 | S | 0 | 0 | A |
| $\bar{3}_z m'_x$ | 0 | 0 | S | 0 | 0 | A |
| $4_z/m_z$ | 0 | 0 | S | 0 | 0 | A |
| $4_z/m_z m'_x m'_{xy}$ | 0 | 0 | S | 0 | 0 | A |
| $\bar{4}_z$ | 0 | 0 | S | 0 | 0 | A |
| $\bar{4}_z 2'_x m'_{xy}$ | 0 | 0 | S | 0 | 0 | A |
| $6_z/m_z$ | 0 | 0 | S | 0 | 0 | A |
| $6_z/m_z m'_x m'_y$ | 0 | 0 | S | 0 | 0 | A |
| $\bar{3}_z$ | 0 | 0 | S | 0 | 0 | A |

## 3. Conclusions

Thus, the magnetic point groups allow determining kind of distributions of electrical polarization in magnetization distribution heterogeneities appeared on crystal lattice defect. Spontaneous polarization produced by the inhomogeneous magnetoelectric effect can be realized in crystal with arbitrary symmetry. The zero degree domain walls with non-polar magnetic point group have odd electrical polarization distribution. Any zero degree domain walls have coupled electric charge in their volume. Thus, their magnetoelectric properties can be detected using the homogeneous electric field. All zero degree domain walls with time-noninvariant chirality have identical type of spatial distribution of the magnetization and polarization.